\documentclass[conference]{IEEEtran}
\IEEEoverridecommandlockouts

\usepackage{amsmath,amssymb,amsfonts}
\usepackage{algorithmic}
\usepackage{graphicx}
\usepackage{textcomp}
\usepackage{booktabs}
\usepackage{multirow,siunitx}
\usepackage{url}
\usepackage{xcolor}
\usepackage[numbers]{natbib}
\usepackage[justification=raggedright]{caption}
\def\BibTeX{{\rm B\kern-.05em{\sc i\kern-.025em b}\kern-.08em
    T\kern-.1667em\lower.7ex\hbox{E}\kern-.125emX}}
\begin{document}

\title{Everyone-Can-Sing: Zero-Shot Singing Voice Synthesis and Conversion with Speech Reference\\}

\author{\IEEEauthorblockN{Shuqi Dai}
\IEEEauthorblockA{\textit{Carnegie Mellon University}}

\and
\IEEEauthorblockN{Yunyun Wang}
\IEEEauthorblockA{\textit{Princeton University}} 
\and
\IEEEauthorblockN{Roger B. Dannenberg}
\IEEEauthorblockA{\textit{Carnegie Mellon University}}
\and
\IEEEauthorblockN{Zeyu Jin}
\IEEEauthorblockA{\textit{Adobe Research}}
}

\maketitle

\begin{abstract}

We propose a unified framework for Singing Voice Synthesis (SVS) and Conversion (SVC), addressing the limitations of existing approaches in cross-domain SVS/SVC, poor output musicality, and scarcity of singing data. 
Our framework enables control over multiple aspects, including language content based on lyrics, performance attributes based on a musical score, singing style and vocal techniques based on a selector, and voice identity based on a speech sample. 
The proposed zero-shot learning paradigm consists of one SVS model and two SVC models, utilizing pre-trained content embeddings and a diffusion-based generator. The proposed framework is also trained on mixed datasets comprising both singing and speech audio, allowing singing voice cloning based on speech reference. 
Experiments show substantial improvements in timbre similarity and musicality over state-of-the-art baselines, providing insights into other low-data music tasks such as instrumental style transfer. Examples can be found at: \texttt{\url{everyone-can-sing.github.io}}.

\end{abstract}


\section{Introduction}
Singing voice synthesis (SVS), which generates singing voice signals from music scores, is gaining increasing importance in generative AI and benefiting various applications in music production and entertainment.  Recent advances in deep-learning-based audio synthesis, such as acoustic models \cite{ren2020fastspeech}, neural vocoders~\cite{kong2020hifi, lee2022bigvgan}, and tokenizer-based codec models~\cite{kumar2024high, zeghidour2021soundstream}, have greatly improved models' ability to reproduce singing voices from training data ~\cite{he-etal-2023-rmssinger, lu2020xiaoicesing, huang2023singing}. However, despite these advances, current SVS models struggle with unseen voices in zero-shot settings, especially when the voice reference is very brief (just a few seconds). This challenge is exacerbated by the dearth of relevant data for learning musicality, expressiveness, and other intricacies that define singing. The vast disparity between available speech data and its limitations as a proxy for singing further compounds this issue. Consequently, zero-shot SVS performance remains subpar, especially in musicality, acoustic quality, and voice similarity. While recent progress has been made in zero-shot speech synthesis through voice-content disentanglement ~\cite{wang2024gr0, choi2021neural, wang2023neural, choi2022nansy++}, the field still awaits a comprehensive solution to overcome these limitations.
 
Zero-Shot Singing Voice Conversion (SVC) faces similar challenges. SVC alters a singer’s voice while preserving the song’s content. Unlike SVS, it converts from an existing singing sample rather than using a symbolic (score) as input. In zero-shot settings, current approaches rely on long voice sample references \cite{so-vits-svc} and are commonly less effective with cross-domain speech targets ~\cite{10447981, 10389711}. While zero-shot speech conversion models can function with shorter voice samples \cite{choi2021neural, wang2023neural}, they fail to address and capture the essence of singing, particularly expressiveness and musicality. Moreover, speech and singing timbre of the same person can differ significantly; the input singing pitch range may vary widely from the short target speech; singing techniques such as vibrato, head voice, and high notes are often ignored by speech conversion models.

This paper addresses the challenges of zero-shot cross-domain SVS and SVC, using only a 5-second speech audio as a reference. To ground our definition of voice similarity, it is necessary to consider what's being transferred from the voice reference. Past research can be summarized into the following categories: (1) voice timbre transfer~\cite{choi2021neural, wang2024gr0, choi2022nansy++}, (2) voice timbre and prosody transfer ~\cite{wang2023neural}, (3) further incorporating performance attributes such as rhythmic and pitch habits \cite{dai2024expressivesinger} which are often seen as \textit{music performance style transfer} \cite{dai2018music}, or (4) using vague, data-driven definitions ~\cite{10447981, 10389711, wu2022unified}. This paper focuses on (1) voice timbre, while the prosody and performance style are controllable by additional model input conditions. We envision that this design is comparatively more beneficial to real-world applications.

The key to enabling singing style control and voice transfer without changing the content is the disentanglement of these elements in the signal. 
Models such as~\cite{choi2021neural, wang2023neural, shen2023naturalspeech} learn such disentanglement through conditional synthesis with large-scale data training. GR0~\cite{wang2024gr0} uses self-supervised learning to separate voice timbre (global voice embedding) from other components (local content embedding). Our model is built on the assumption that voice, content, and singing styles are naturally disentangled. Our SVC model utilizes pre-trained disentangled representations such as speaker embedding and content embedding in GR0. To achieve good musicality, our SVS model is conditioned on more granular representations called \textit{expressive performance attributes} including pitch curves, loudness contour, and pronunciation, while adding controls for language, singing style, and technique, following the framework of ExpressiveSinger~\cite{dai2024expressivesinger}. 

Finally, we address singing data scarcity by utilizing abundant speech data in the scheduled training framework, incorporating pre-training, fine-tuning, and mixed-training strategies.

\begin{figure}
    \vspace{-0.4cm}
    \centering                                              
    \includegraphics[width=0.48\textwidth]{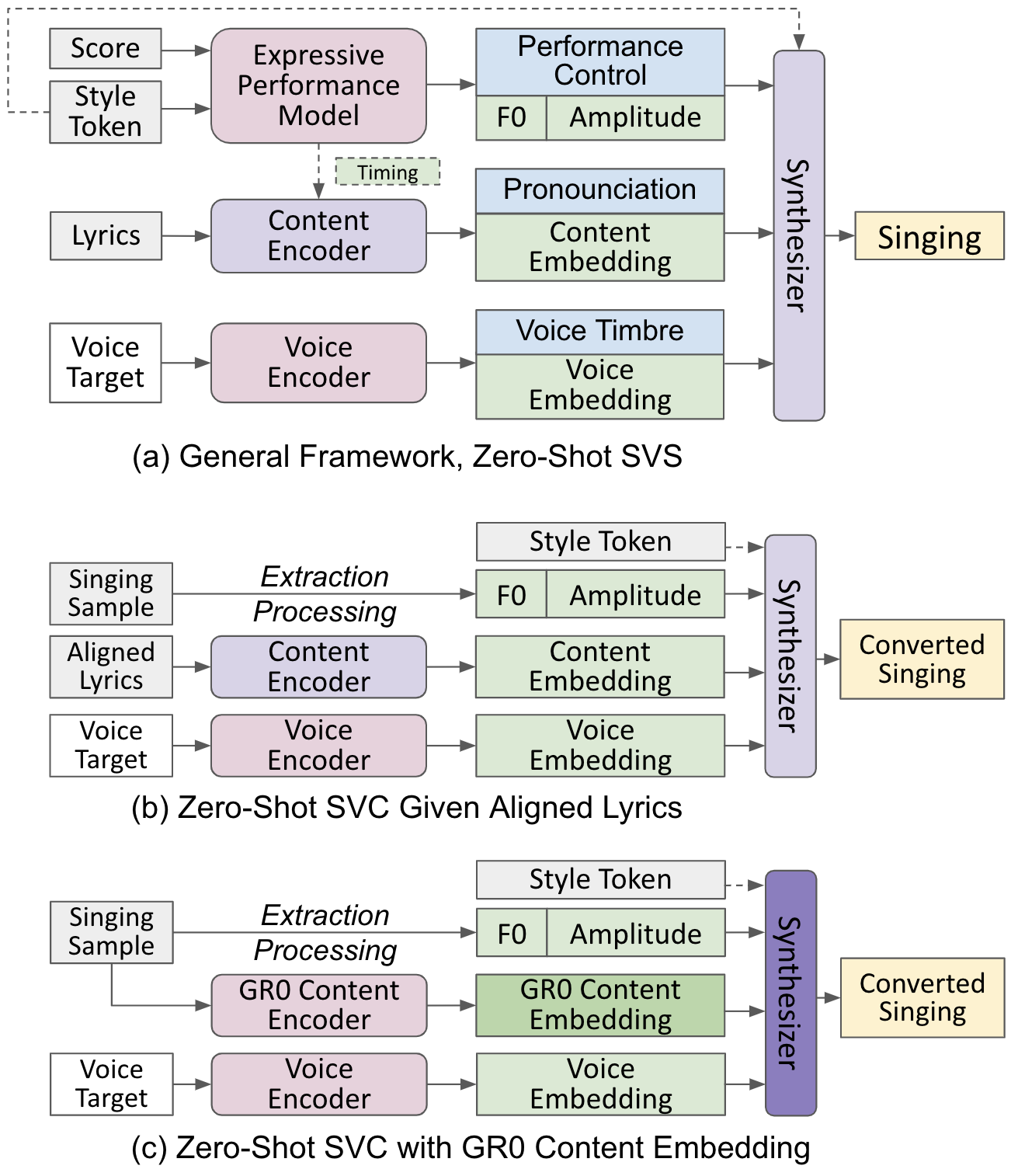}
    \caption{Proposed zero-shot SVS and SVC framework.}
    \label{fig:framework}
    \vspace{-0.6cm}
\end{figure}

Our unified framework, illustrated in Fig.\ref{fig:framework}, introduces three models: one for zero-shot SVS and two for zero-shot SVC. First, building on ExpressiveSinger and inspired by GR0’s self-supervised training, we develop a zero-shot SVS with improved timbre embedding and a mixed training approach using both singing and speech data. We then extend this framework to zero-shot SVC with lyrics annotation for language, style, and technique control. Finally, we incorporate GR0’s local content embedding, enabling conversion without additional lyrics input. The key contributions of this paper are:

\begin{itemize}
    \item [(1)] We design a system that simultaneously integrates high-quality SVS, SVC, and zero-shot capabilities for cross-domain scenarios using only a 5-second speech reference.
    \item [(2)] Our framework provides fine-grained control over: linguistic content (via lyrics), performance attributes (via musical score), singing style and vocal techniques (via a selector), and voice identity (via speech samples) in zero-shot scenarios, significantly advancing the flexibility and expressiveness of current zero-shot SVS and SVC.
    \item [(3)] We emphasize the crucial role of performance attributes, and demonstrate that better disentanglement and control of singing components—especially expressive performance attributes—improve singing generation quality and mitigate data scarcity.
    \item [(4)] We introduce a method for integrating speech data into singing model training without compromising output quality, addressing the scarcity of singing data, and offering insights for similar low-data scenarios.
    \item [(5)] Experiments including extensive ablation studies and large-scale subjective evaluations, show improvements in singing performance and also offer insights for speech.
\end{itemize}

\section{Method}
\subsection{Unified Framework}
Our method is built on a unified framework, as shown in Fig.\ref{fig:framework}, from which we derive three models: one zero-shot SVS model and two zero-shot SVC models. Inspired by ExpressiveSinger~\cite{dai2024expressivesinger}, the core principle here is to break down singing into three components: musical perspective attributes (\textit{expressive performance control}\cite{dai2018music}), pronunciation related to lyrics, and voice timbre. Each component is modeled separately, and the synthesis process learns to disentangle them in training. During inference, the main synthesizer takes all three components and generates the output singing. It consists of two parts: an acoustic model that generates a mel-spectrogram from the component outputs and a neural vocoder that converts the mel-spectrogram into a waveform.

The three components are defined as follows. First, expressive performance control attributes capture how a singer interprets a music score with personal style and emotion. Building on ExpressiveSinger, we utilize two key attributes: Fundamental Frequency (F0) curves for pitch contour and amplitude envelopes for dynamics, with performance timing embedded in their respective time frames. The second component, lyrics pronunciation, is represented by a time sequence of embedding vectors that also includes phoneme timing. The third component, voice timbre, is a time-independent embedding capturing timbre information from speech or singing audio. Although performance control captures singing style, lyrics pronunciation and timbre can also vary with style (e.g., the same speaker’s timbre may change). To address this, we add a style token as a conditional input. 

We tailor each component to the specific task. In SVS, performance control attributes are generated from the musical score and style tokens, while in SVC, they are extracted from the input singing sample via signal processing. In SVC, lyrics pronunciation is either disentangled from the singing sample or generated using aligned lyrics. For zero-shot voice timbre modeling, the voice embedding is easily swapped as needed.

While different modules employ various training methods, the acoustic model of the main synthesizer consistently uses a diffusion process to generate mel-spectrograms as the first step. Our training strategy involves pre-training, fine-tuning, and mixed-training, leveraging both singing and speech data.

\subsection{Zero-Shot Singing Voice Synthesis}
As shown in Fig.\ref{fig:framework}(a), our zero-shot SVS model takes as input a symbolic score (with pitch and duration for each note), a style token indicating genre and technique, lyrics aligned with the score, and a 5-second speech reference. The output is a singing audio waveform that matches the timbre of the speech reference while adhering to the score and style controls. The model architecture largely follows ExpressiveSinger, with two key modifications: (1) using Resemblyzer, a pre-trained voice encoder, instead of a symbolic Singer ID, and (2) adding a Leaky ReLU activation layer to the pronunciation content encoder after the transformer and fully connected layers.

We directly adapt ExpressiveSinger’s trained modules for performance timing, F0, and amplitude control. The generated performance timing feeds both the F0 and amplitude modules and aligns phonemes in the content encoder. We use Resemblyzer~\cite{resemblyzer} as the pre-trained speaker embedding model and BigVGAN~\cite{lee2022bigvgan}, trained on general audio, as the main synthesizer vocoder. Thus, only the pronunciation content encoder and the main acoustic model require training, which follows a diffusion-based procedure \cite{engel2019ddsp} with the same reconstruction losses and settings as ExpressiveSinger. We employ mixed training at a 1:1 ratio of singing and speech data, randomly sampling from each domain dataset. During training, F0 and amplitude are extracted from ground truth (GT) singing or speech samples, as is the voice target embedding. We also adjust the singing F0 and amplitude extraction algorithm \cite{dai2024expressivesinger} to better accommodate speech.

During inference, we replace the voice target with the unseen speech reference. To handle potential pitch range mismatches between speech and singing, we offer a pitch adjustment method that shifts the target music score into a range within one octave of the speech reference.

\subsection{Zero-Shot SVC Given Lyrics Alignment}
Fig.\ref{fig:framework}(b) takes a singing sample, an unseen speech reference, and phoneme-level lyrics (aligned with the singing sample) as input to the pronunciation content encoder, producing the converted singing. Although aligned lyrics can be extracted via recognition and alignment models, their low accuracy necessitates an annotated dataset for this step.

This model is a minor modification of the SVS model. During inference, the F0 curves and amplitude envelopes are directly extracted from the singing sample rather than generated by the model. Since this change does not affect training, we can reuse the trained SVS modules without further adjustments.

\subsection{Zero-Shot SVC With Local Content Embedding}
Fig.\ref{fig:framework}(c) requires only a singing sample and an unseen speech reference to generate the converted singing; style tokens are optional. Here, the pronunciation embedding is extracted directly from the singing sample via GR0’s content encoder~\cite{wang2024gr0}, removing the need for aligned lyrics. Adapted from a pre-trained wav2vec 2.0 model with CTC loss~\cite{baevski2020wav2vec}, this encoder does not include voice timbre information~\cite{wang2024gr0}.

We train a new acoustic model for the synthesizer by replacing the SVS lyrics content encoder with the GR0 embedding, keeping all other training settings except the training. Since the GR0 content encoder has already been trained on extensive speech datasets, it effectively disentangles the pronunciation component, allowing us to train on singing data alone without mixed training. During inference, we apply the same pitch adjustments as in the previous models.

\section{Experiments}
\subsection{Experiment Settings}
We use different datasets to train the modules in our framework. Specifically, all three models’ acoustic synthesizers were trained on the same singing data as ExpressiveSinger, comprising $62$ hours of recordings from $50$ singers in three languages and multiple style labels \cite{dai2024expressivesinger, singstyle111}, with identical training, testing, and validation splits. 
For the zero-shot SVS model (Fig.\ref{fig:framework}(a)) and the zero-shot SVC model with aligned lyrics (Fig.\ref{fig:framework}(b)), we employ mixed training on both singing and speech data. The speech data is from the LibriTTS-R dataset \cite{koizumi2023libritts}, which spans $585$ hours of English speech from $2,456$ speakers. Data representation and processing follow ExpressiveSinger, with minor adjustments to F0 and amplitude extraction for speech. Input and output audio is standardized at $22.05$ kHz, while mel-spectrograms, F0, amplitude, and the vocoder use $256$ hop size, $1024$ window size, $1024$ FFT size, $80$ mel bins, and a frequency range of $0$ to $11,025$ Hz.

We reuse ExpressiveSinger’s pre-trained expressive performance modules for our SVS model and adapt Resemblyzer’s speech-trained model as the voice encoder. GR0’s content encoder is also trained on speech, while BigVGAN, pre-trained on general audio, is fine-tuned on singing data \cite{singstyle111}. The acoustic model of the main synthesizer is trained for $900k$ iterations on four NVIDIA A100 GPUs, with a batch size of $32$ and a learning rate of $2e-4$. We set the diffusion steps to \( T = 1000 \) using a linear noise schedule \(\beta\) from $0.0001$ to $0.02$, and the diffusion step embeddings use $128$, $512$, and $512$ channels for the input, middle, and final layers, respectively. During inference, we employ DDIM fast sampling \cite{song2020denoising}.

\subsection{Subjective Evaluation}
Given the lack of reliable objective evaluation methods for singing, particularly for cross-domain similarity between singing and speech, we conduct two subjective evaluations: a comparison study to benchmark our models against baselines, and an ablation study to validate our design, control, and generalization. In this experiment, we select $50$ distinct voice audios as zero-shot target references, each $5$--$7$ seconds in length, evenly distributed across speech and singing, male and female voices, various timbres and pitch ranges, age groups, and languages. We employ \textit{pop} and \textit{opera} as style conditions, limiting vocal techniques to \textit{normal} and \textit{vibrato}~\cite{singstyle111}.

We establish a range of ablation conditions to generate singing and evaluate model performance, encompassing variations in models, training data, reference audio type (singing or speech), cross-language or cross-gender scenarios, pitch range adjustments, and style control types. We also include baseline models, human singing, and mismatched reference targets. Each condition provided at least $80$ demos (except for one baseline \cite{wu2022unified} where only four demo pairs are available), resulting in a total of $1538$ demos.

Each subjective evaluation survey contains $10$ singing pairs, each comprising a test singing and its paired reference target. The pairs are randomly selected from different conditions to ensure an even distribution across all types and demos. For each pair, participants rate two Mean Opinion Scores (MOS) on a 1–5 scale: (1) the quality of the test singing (considering musicality, pronunciation, pitch accuracy, naturalness, and expressiveness), and (2) the voice (timbre) similarity between the test singing and the reference. A validation test with an extremely low-quality singing sample (covered in white noise) is included to filter out careless ratings. In total, 503 surveys were collected, and 487 remain valid after validation.

\subsubsection{Comparison Study}
\begin{table}[t]
  \begin{tabular}{cccc}
    \toprule
    Exp. & Target Reference & MOS & SIM-MOS\\
    \midrule
    Human & Singing & 4.43 {$\pm$} 0.13 & 4.08 {$\pm$} 0.16 \\
    Mismatch & Singing/Speech & / & 3.01 {$\pm$} 0.16 \\
    ExpressiveSinger\cite{dai2024expressivesinger} & Singing & 4.29 {$\pm$} 0.14 &  4.24 {$\pm$} 0.16 \\
    \midrule
    Zero-shot SVS(a) \textit{Ours} & Speech & 3.98 {$\pm$} 0.07 & \textbf{3.78 {$\pm$} 0.12} \\
    \midrule
    Unified\cite{wu2022unified} & Singing & 3.33 {$\pm$} 0.23 & 3.27 {$\pm$ 0.18} \\
    GR0\cite{wang2024gr0}& Speech & 2.92 {$\pm$} 0.15 & 3.51 {$\pm$} 0.15\\
    \midrule
    Zero-shot SVC(b) \textit{Ours} & Speech & 3.95 {$\pm$} 0.07 & 3.61 {$\pm$} 0.11  \\
    Zero-shot SVC(c) \textit{Ours} & Speech & \textbf{4.0 {$\pm$} 0.08} & 3.59 {$\pm$} 0.12\\
  \bottomrule
\end{tabular}
\caption{Synthesized singing quality (MOS) and similarity (SIM-MOS) comparison against baselines. ``Human'' refers to actual human singing samples. ``Zero-shot SVC(b)'' represents our zero-shot SVC model with aligned lyrics, while ``Zero-shot SVC(c)'' is the zero-shot SVC with GR0 content embedding. ``Target Reference'' indicates whether the paired reference audio is a speech or singing sample. All scores are reported with a 95\% confidence interval with significance levels $<$ 0.001.}
\label{tab:exp}
\vspace{-0.3cm}
\end{table}

We compare eight different models and conditions, summarized in Table~\ref{tab:exp}. All three proposed models (``\textit{Ours}'') use speech audio as target references with pitch adjustment; mixed training is applied to all but the SVC model with GR0 content embedding. As baselines, we include human singing as ground truth, paired with a different singing from the same singer as the target reference. We also add a mismatch condition where a random singing demo is paired with a non-matching singing or speech reference. Because no accurate open-source zero-shot SVS model is available, we use ExpressiveSinger, the state-of-the-art SVS model without zero-shot capability, as a baseline. Its outputs are paired with the same singer’s human singing sample, resulting in a very high similarity MOS (SIM-MOS). For SVC, we use the public demo from an existing zero-shot SVC model~\cite{wu2022unified} and, since no other zero-shot SVC models are available, we also include a GR0 model trained on speech as another baseline.

As shown in Table~\ref{tab:exp}, ExpressiveSinger achieves the highest scores for both singing quality and reference similarity; however, it is not a zero-shot model, as its reference is drawn from a singer in the training data, resulting in near ground-truth SIM-MOS. While our zero-shot SVS model slightly lags ExpressiveSinger in singing quality, it delivers the highest similarity among all zero-shot approaches, including conversion models. Both proposed SVC models also significantly outperform the SVC baselines in terms of singing quality and timbre similarity. Among our three proposed models, singing quality remains consistently high.

\newcommand\ChangeRT[1]{\noalign{\hrule height #1}}

\subsubsection{Ablation Study}
\begin{table}[t]
  \resizebox{\linewidth}{!}{
  \begin{tabular}{c!{\vrule width 1.5pt}cccccc}
    \ChangeRT{1.8pt}
    & \multicolumn{2}{c!{\vrule width 1.5pt}}{\textbf{Zero-shot SVS}} & \multicolumn{2}{c!{\vrule width 1.5pt}}{\textbf{Zero-shot SVC(b)}} & \multicolumn{2}{c}{\textbf{Zero-shot SVC(c)}} \\
    \hline
     \textbf{Exp.} &  \multicolumn{1}{c|}{\textbf{MOS}} &  \multicolumn{1}{|c!{\vrule width 1.5pt}}{\textbf{SIM-MOS}} & \multicolumn{1}{c|}{\textbf{MOS}} &  \multicolumn{1}{|c!{\vrule width 1.5pt}}{\textbf{SIM-MOS}} & \multicolumn{1}{c|}{\textbf{MOS}} &  \multicolumn{1}{|c}{\textbf{SIM-MOS}} \\
    \ChangeRT{1.2pt}
    Speech Ref. & \multicolumn{1}{c|}{3.9 {$\pm$} 0.13} & \multicolumn{1}{|c!{\vrule width 1.5pt}}{3.5 {$\pm$} 0.18} & \multicolumn{1}{c}{3.95 {$\pm$} 0.05} & \multicolumn{1}{|c!{\vrule width 1.5pt}}{3.47 {$\pm$} 0.1} & \multicolumn{1}{c}{3.93 {$\pm$} 0.11} & \multicolumn{1}{|c}{3.5 {$\pm$} 0.14} \\
    Singing Ref. & \multicolumn{1}{c|}{\bf 4.03 {$\pm$} 0.09} & \multicolumn{1}{|c!{\vrule width 1.5pt}}{\bf 3.84 {$\pm$} 0.11} & \multicolumn{1}{c}{3.95 {$\pm$} 0.08} & \multicolumn{1}{|c!{\vrule width 1.5pt}}{\bf 3.66 {$\pm$} 0.11} & \multicolumn{1}{c}{\bf 4.11 {$\pm$} 0.12} & \multicolumn{1}{|c}{\bf 3.76 {$\pm$} 0.16} \\
    \hline
    Mix Train. & \multicolumn{1}{c|}{\bf 4.01 $\pm$ 0.08} & \multicolumn{1}{|c!{\vrule width 1.5pt}}{\bf 3.8 $\pm$ 0.13} & \multicolumn{1}{c}{\bf 3.95 {$\pm$} 0.05} & \multicolumn{1}{|c!{\vrule width 1.5pt}}{\bf 3.59 $\pm$ 0.07 } & \multicolumn{1}{c}{/} & \multicolumn{1}{|c}{/} \\
    Singing Train. & \multicolumn{1}{c|}{3.83 $\pm$ 0.21} & \multicolumn{1}{|c!{\vrule width 1.5pt}}{3.46 $\pm$ 0.49} & \multicolumn{1}{c}{3.76 {$\pm$} 0.35} & \multicolumn{1}{|c!{\vrule width 1.5pt}}{3.47 {$\pm$} 0.53} & \multicolumn{1}{c}{4.0 {$\pm$} 0.08} & \multicolumn{1}{|c}{3.61 $\pm$ 0.1} \\
    \hline
    Adjust Pitch & \multicolumn{1}{c|}{3.97 {$\pm$} 0.09} & \multicolumn{1}{|c!{\vrule width 1.5pt}}{\bf 3.78 {$\pm$} 0.12} & \multicolumn{1}{c}{3.96 {$\pm$} 0.08} & \multicolumn{1}{|c!{\vrule width 1.5pt}}{\bf 3.61 {$\pm$} 0.11} & \multicolumn{1}{c}{3.98 {$\pm$} 0.09} & \multicolumn{1}{|c}{\bf 3.59 {$\pm$} 0.12} \\
    Original Pitch & \multicolumn{1}{c|}{\bf 4.02 {$\pm$} 0.12} & \multicolumn{1}{|c!{\vrule width 1.5pt}}{ 3.6 {$\pm$} 0.18} & \multicolumn{1}{c}{\bf 4.03 {$\pm$} 0.07} & \multicolumn{1}{|c!{\vrule width 1.5pt}}{3.5 {$\pm$} 0.1} & \multicolumn{1}{c}{\bf 4.09 {$\pm$} 0.17} & \multicolumn{1}{|c}{3.58 {$\pm$} 0.23} \\
    \hline
    Cross Language & \multicolumn{1}{c|}{\bf 3.99 {$\pm$} 0.09} & \multicolumn{1}{|c!{\vrule width 1.5pt}}{\bf 3.82 {$\pm$} 0.12} & \multicolumn{1}{c}{3.88 {$\pm$} 0.08} & \multicolumn{1}{|c!{\vrule width 1.5pt}}{3.58 {$\pm$} 0.11} & \multicolumn{1}{c}{4.0 {$\pm$} 0.09} & \multicolumn{1}{|c}{\bf 3.63 {$\pm$} 0.11} \\
    Same Language & \multicolumn{1}{c|}{3.97 {$\pm$} 0.12} & \multicolumn{1}{|c!{\vrule width 1.5pt}}{3.56 {$\pm$} 0.17} & \multicolumn{1}{c}{\bf 4.02 {$\pm$} 0.07} & \multicolumn{1}{|c!{\vrule width 1.5pt}}{\bf  3.6 {$\pm$} 0.1} & \multicolumn{1}{c}{\bf 4.14 {$\pm$} 0.31} & \multicolumn{1}{|c}{ 3.45 {$\pm$} 0.41} \\
    \hline
    Diff. Gender  & \multicolumn{1}{c|}{\bf 4.01 {$\pm$} 0.09} & \multicolumn{1}{|c!{\vrule width 1.5pt}}{3.67 {$\pm$} 0.12} & \multicolumn{1}{c}{3.88 {$\pm$} 0.07} & \multicolumn{1}{|c!{\vrule width 1.5pt}}{\bf 3.9 {$\pm$} 0.11} & \multicolumn{1}{c}{3.76 {$\pm$} 0.14} & \multicolumn{1}{|c}{\bf 3.71 {$\pm$} 0.16} \\
    Same Gender & \multicolumn{1}{c|}{3.93 {$\pm$} 0.13} & \multicolumn{1}{|c!{\vrule width 1.5pt}}{\bf  3.81 {$\pm$} 0.17} & \multicolumn{1}{c}{\bf 4.03 {$\pm$} 0.07} & \multicolumn{1}{|c!{\vrule width 1.5pt}}{ 3.61 {$\pm$} 0.1} & \multicolumn{1}{c}{\bf 4.2 {$\pm$} 0.1} & \multicolumn{1}{|c}{3.54 {$\pm$} 0.14} \\
    \hline
    Pop Style & \multicolumn{1}{c|}{\bf 3.98 {$\pm$} 0.07} & \multicolumn{1}{|c!{\vrule width 1.5pt}}{\bf 3.72 {$\pm$} 0.1} & \multicolumn{1}{c}{\bf 3.98 {$\pm$} 0.06} & \multicolumn{1}{|c!{\vrule width 1.5pt}}{\bf 3.57 {$\pm$} 0.08} & \multicolumn{1}{c}{\bf 4.0 {$\pm$} 0.08} & \multicolumn{1}{|c}{\bf  3.61 {$\pm$} 0.11} \\
    Opera Style & \multicolumn{1}{c|}{3.85 $\pm$ 0.15} & \multicolumn{1}{|c!{\vrule width 1.5pt}}{3.42 $\pm$ 0.23} & \multicolumn{1}{c}{3.8 {$\pm$} 0.13} & \multicolumn{1}{|c!{\vrule width 1.5pt}}{3.45 {$\pm$} 0.21} & \multicolumn{1}{c}{3.81 $\pm$ 0.17} & \multicolumn{1}{|c}{3.4 $\pm$ 0.19} \\
  \ChangeRT{1.8pt}
\end{tabular}
}
\caption{Ablation study on singing quality and timbre similarity scores for our three models. ``Ref." refers to the target reference audio. ``Mix Train." indicates mixed training, while ``Singing Train." refers to using only singing data. ``Adjust Pitch" applies pitch adjustment during inference, while "Original Pitch" retains the source or score input pitch. ``Cross Language" refers to the target reference and source (or input singing lyrics) being in different languages. ``Diff Gender" indicates the target reference and source (or score pitch range) are of different genders.}
\label{tab:exp2}
\vspace{-0.3cm}
\end{table}

As shown in Table~\ref{tab:exp2}, we test various ablation conditions for our models, which differ from those in Table~\ref{tab:exp}. For instance, the first group with speech and singing references includes models both with and without pitch adjustment, whereas Table~\ref{tab:exp} always uses pitch adjustment.

When the target reference is singing, the generated singing quality remains similar to that using speech references (except for a notable improvement in zero-shot SVC(c)), indicating that our models consistently produce high-quality singing regardless of reference type. However, voice timbre similarity improves with singing references, likely because the difference between speaking and singing voices can be substantial, making intra-domain comparisons more favorable.

The mixed training strategy enhances both quality and similarity across all models. In contrast, training only on singing data leads to higher variance in these scores, likely due to the limited data scale in zero-shot settings, which could cause instability. Pitch adjustment aligns source and target pitch ranges, boosting similarity but sometimes reducing singing quality; for example, lowering a song’s pitch to match the normal speaking range can make it sound unnaturally low.

Mismatch between source and target languages does not degrade the quality and may even increase similarity, as cross-language scenarios can make timbre differences harder to judge due to varying pronunciation and prosody. When source and target genders differ, SVS quality remains mostly unaffected, although similarity declines slightly. However, for the two SVC models, quality drops significantly while similarity improves, likely due to larger pitch shifts in cross-gender conversion. Finally, our models perform better on pop than on opera, as opera imposes higher technical demands and the gap between speaking and opera singing is typically greater.

\section{Conclusion}
We introduced a unified framework for zero-shot SVS and SVC, addressing challenges in cross-domain voice generation, musicality, controllability, and data scarcity. Our approach integrates different singing component embeddings with a diffusion-based synthesizer, leveraging both singing and speech data. By offering fine-grained control over lyrics, performance attributes, singing styles, and vocal timbre, it achieves high-quality singing with significant improvements in both timbre similarity and musicality compared to state-of-the-art baselines. Experiments validate the effectiveness of our models, offering valuable insights for future music generation tasks such as instrumental timbre style transfer.

\newpage
\bibliographystyle{unsrt}
\bibliography{sing}

\end{document}